# Journal of Materials Chemistry A

RSCPublishing

## ARTICLE

## $N_2H$: A Novel Polymeric Hydronitrogen as a High Energy Density Material

Ketao Yin[a], Yanchao Wang[a,*], Hanyu Liu[a], Feng Peng[a] and Lijun Zhang[b,*]

**ABSTRACT:** The polymeric phase of nitrogen connected by the lower (than three) order N-N bonds has been long sought after for the potential application as high energy density materials. Here we report a hitherto unknown polymeric $N_2H$ phase discovered in the high-pressure hydronitrogen system by first-principle structure search method based on particle swarm optimization algorithm. This polymeric hydronitrogen consists of quasi-one-dimensional infinite armchair-like polymeric N chains, where H atoms bond with two adjacent N located at one side of armchair edge. It is energetically stable against decomposition above ~33 GPa, and shows novel metallic feature as the result of pressure-enhanced charge transfer and delocalization of π electrons within the infinite nitrogen chains. The high energy density (~4.40 KJ/g), high nitrogen content (96.6%), as well as relatively low stabilization pressure, make it a possible candidate for high energy density applications. It also has lattice dynamical stability down to the ambient pressure, allowing for the possibility of kinetic stability with respect to variations of external conditions. Experimental synthesis of this novel phase is called for.

## I. Introduction

Due to the remarkable difference in average bond energy between the N-N single-bond/double-bond (160/418 kJ/mol) and the triple-bond of $N_2$ molecule (954 kJ/mol),[1] the transformation from the non-molecular phase to the molecular state essentially accompanies with a large amount of energy released, thus the non-molecular phase of "polymeric nitrogen" with low-order N-N bonds has been long-standing desired as a high energy density material (HEDM) for propellant and explosive applications.[2,3] One of effective approaches to stabilize such polymeric nitrogen phases is to modify the chemical bonding behaviour of nitrogen by the application of pressure. Along this direction, a single-bond phase of solid nitrogen with the theoretically predicted cubic gauche (cg-N) structure[4] has been experimentally synthesized by Eremets et al. at high pressure and high temperature (110 GPa, 2000 K).[5,6] The cg-N is predicted to have a more than three times higher energy storage capacity than the most powerful energetic materials.[6–8] Moreover it is a clean energy source as the final product being environmentally harmless nitrogen gas. However, the ultrahigh pressure required to stabilize it (~94 GPa[5] and 127 GPa at room temperature[9]) has precluded its practical applications.

To lower the stabilization pressure, considerable attention has been focused on compressing nitrogen-containing compounds with other elements, including the alkali-metal azides, $AN_3$ (A = Li, Na, K, Cs),[9–12] the CNO system (CO/$N_2$ mixture),[13] the CN system ($C_3N_{12}$),[14] etc. Particularly, by using the mixture of $NaN_3$ and $N_2$ as the precursor, it has been observed experimentally that the cg-N can be formed at the lower pressure (50 GPa) than that in the pure $N_2$.[9] Raza et al.[13] predicted via first-principle calculations that the CNO system is transformed to the structure having a three-dimensional bonded framework at 52 GPa. The mechanism underlying the reduced stabilization pressure might be attributed to the chemical precompression resulted from other elements,[15] existence of more types of N-N bonding than the $N_2$ molecule,[16] or assistance of additional bondings between N and other elements.[13]

Hydronitrogen represents a rich family of nitrogen-containing materials, of which the extremely high mass ratio of nitrogen makes it particularly attractive for HEDM. At ambient condition, the ground-state hydronitrogen is well known to adopt the ammonia phase consisting of $NH_3$ molecules. Under specific conditions some metastable phases can be synthesized, including hydrazine ($N_2H_4$),[17] diazene ($N_2H_2$),[18] ammonium azide ($NH_4N_3$),[19] tetrazene ($N_4H_4$)[20] and hydrogen azide ($HN_3$),[21] etc. The diversity of these metastable phases implies the versatile capability of hydrogen in stabilizing a wide range of compounds with the lower order N-N bonds. With increasing pressures, a series of structural phase transitions have been reported, for instance in $NH_3$;[22] moreover, an exotic ionic phase formed by $NH_4^+$ and $NH_2^-$ ions has been theoretically predicted[23] and recently observed in experiments.[24,25] However, all of these phases are composed of isolated molecules, which might transform to the pseudo-non-molecular phases via pressure-induced hydrogen-bond symmetrisation (i.e., molecules connected by the bridged H at the centre),[26] not being the true sense of the polymeric nitrogen phase. It is thus generally accepted that the polymeric nitrogen phase cannot be realized based on hydronitrogen, until 2011 Hu and Zhang[27] proposed based on first-principle calculations that a new hydronitrogen with the stoichiometry of NH (in the space group of $P2_1/m$), containing an infinite nitrogen zigzag chain passivated sideward by H atoms, can be stabilized above moderate pressure (~36 GPa). But the subsequent experiments found that the molecular $NH_4N_3$ phase preserves up to ~70GPa,[28,29] clearly not supporting the prediction. Quite recently, Wang et al. reported the formation of a mixture of nitrogen backbone oligomers — finite chains of the low-order N-N bonds (terminated by H) — via direct reaction between $H_2$ and $N_2$ at room temperature and high pressure of ~35 GPa.[30] Though the structures of new phases not completely resolved yet,[30] this work underlies a promising perspective of synthesizing rich hydronitrogen phases containing the low-order N-N bonds at high pressures.

With the aim of getting more insights into the open question whether the polymeric nitrogen phase can exist in hydronitrogens or other nitrogen-containing systems, we herein perform a comprehensive crystal structure search for energetically stable hydronitrogen phases with varied H and N contents up to 200 GPa by first-principle density functional theory (DFT) total energy calculations guided by an in-house developed Crystal structure AnaLYsis by Particle Swarm Optimization (CALYPSO) methodology.[31,32] A previously unknown polymeric $N_2H$ phase of hydronitrogen is discovered, which is energetically stable against decomposition above a relatively low pressure of ~33 GPa. This phase consists of polymeric framework of infinite armchair-like nitrogen chains, and shows novel metallic feature caused by the charge transfer process and delocalization of π electrons within the quasi-one-dimensional nitrogen chains. As expected it exhibits a high energy density of ~4.40 KJ/g, originating from the alternate N-N single-bonds and double-bonds existing in the infinite nitrogen chains.

## II. Computational approaches

Our variable-composition crystal structure prediction is based on a global minimum search of the free energy surfaces calculated by *ab initio* DFT total-energy calculations, through the particle swarm optimization algorithm implemented in the CALYPSO code.[33–35] The key feature of this methodology is its capability of predicting the ground-state stable structure of materials with only the knowledge of chemical composition at given external conditions (for example, pressure), not relying on any prior known structural information. Its validity in crystal structure prediction has been robustly demonstrated by its successful applications in a variety of material systems, ranging from elements to binary and ternary compounds.[36–41] See Supplementary Methods for more details on the crystal structure search procedures. The total-energy calculations and local structure optimization are carried out by using the plane wave basis, projected augmented wave (PAW) potentials,[42] and generalized gradient approximation with the Perdew–Burke–Ernzerhof exchange-correlation functional[43] as implemented in the Vienna ab initio simulation package.[44] The frozen-core all-electron PAW potentials have been used, with $2s^22p^3$ (cutoff radius 1.5 a.u.)

and $1s^1$ (cutoff radius 1.1 a.u.) treated as valence electrons for N and H, respectively. During structure prediction steps, an economy set of parameters are employed to accelerate the calculation, following which a high plane-wave cutoff energy (600 eV) and the denser k-mesh sampled in the Brillouin zone (depending on specific structures) have been used to ensure the enthalpies of all the superior structures converged within 1 meV/atom. The van der Waals (vdWs) interaction, which is expected to have important effects on the energetics of molecular systems at the lower pressures, is considered by using the DFT-D2 approach (via adding a semi-empirical dispersion potential to the DFT total energy).[45] Its validity in the current hydronitrogen system has been examined in solid hydrazine ($N_2H_4$) containing the shortest nitrogen chains, where the optimized lattice constants and the nearest inter-molecular distance from the DFT+vdWs approach are (a=3.52 Å, b=5.13 Å, c=4.54 Å) and 3.22 Å, giving overall better agreement with the experimental values, (a=3.56 Å, b=5.78 Å, c=4.53 Å) and 3.19 Å,[17,46] by comparison with the values from the DFT only approach, (a=3.69 Å, b=5.36 Å, c=4.73 Å) and 3.31 Å. The hybrid functional Heyd–Scuseria–Ernzerhof (HSE)[47] is used to achieve an accurate description of the band gap of stable phases. To get a better understanding of chemical bonding behaviors, we employed the Bader charge analysis approach[48] for unbiasedly evaluating the actual charge transfer among N atoms. The phonon dispersion and density of states have been calculated by using a supercell (finite-displacement) approach as implemented in the PHONOPY code[49].

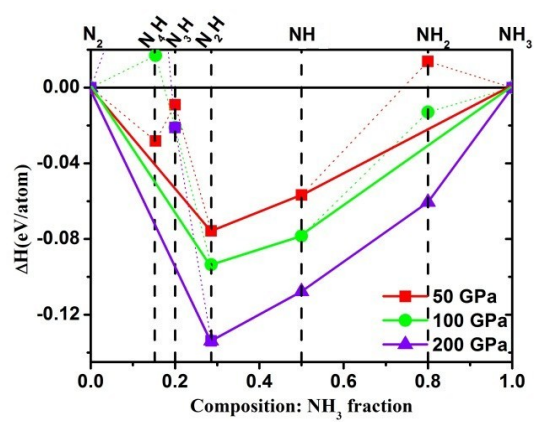

**Figure 1.** Calculated formation enthalpies (ΔH in eV/atom) of various hydronitrogens at different stoichiometries for the chemical reaction between $NH_3$ and $N_2$ at 50 (red), 100 (green) and 200 (violet) GPa. For each $NH_3$ fraction, only ΔH of the most stable phase is shown.

## III. Results and discussion

**Energetic stabilities of nitrogen-rich hydronitrogens at high pressures.** We mainly focus on the hydronitrogens with relatively rich nitrogen content that are more favorable for HEDM. The conventional approach of synthesizing hydronitrogens is by mixing directly $N_2$ and $H_2$.[30,50] Here we propose an alternative routine to synthesize hydronitrogens via the reaction between $N_2$ and $NH_3$ at high pressures. Two advantages are expected from this scenario by comparison with the conventional one: (i) since all the reactants contain nitrogen, the nitrogen-rich phases will be more easily stabilized; (ii) ammonia is more easily under control during the synthesis owing to its relatively high liquefaction and solidification temperature. The energy barrier to be overcome in two reactions might be similar since the predominant contribution is from activating the N-N triple-bonds. With this proposal, from the theoretical side we can also reasonably consider the effect of competing phase ($NH_3$) on phase stabilities of hydronitrogen systems.

We explored the chemical stabilities of a variety of hydronitrogens by calculating their enthalpies of formation at 50, 100, and 200 GPa relative to the products of dissociation into solid $N_2$ and $NH_3$. The results are summarized in the convex hull of the N-H system with respect to $NH_3$ and $N_2$ as the binary variables in Fig. 1 (see Supplementary Fig. S1 for the convex hull containing more unstable structures at 50 GPa). Note that here the results with respect to $NH_3$ and $N_2$ represent merely one selected reference state as a possible decomposition pathway, and we do not perform any calculation on reaction energetics and mechanism, which is beyond the scope of current work. The most stable structures at each composition are obtained by the crystal structure search with up to 8 formula units. As seen, while the extremely nitrogen-rich stoichiometries ($N_4H$ and $N_3H$) cannot be stabilized, the stoichiometry of $N_2H$ is found to be the most stable phase over all the pressure range. The substantial energetic stability of it over other phases (more than 20 meV/atom) implies the advantage of synthesizing it in this reaction. The other stable phases are relatively nitrogen-poor stoichiometries, *i.e.* $NH$ and $NH_2$ (at 200 GPa only), which are actually marginally stable (*i.e.* rather close to the hull) with respect to a mixture of $N_2H$ and $NH_3$.

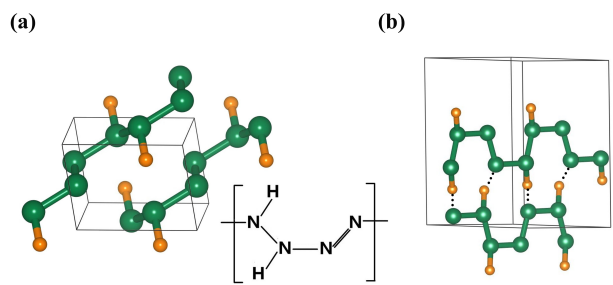

**Figure 2.** Crystal structures and chemical formulations of the $N_2H$ polymeric phase in (a) *P*-1 space group and (b) *P*$2_1$/c space group. Nitrogen and hydrogen atoms are shown in green and orange, respectively. For the *P*$2_1$/c phase, the hydrogen bondings between H and the N belonging to the adjacent chain are denoted as dot lines.

**Ground-state structures of the most stable $N_2H$ stoichiometry.** For the most stable stoichiometry of $N_2H$, all the structures with the lower energies obtained by our structure search consist of an infinite armchair-like polymeric nitrogen chain, where half of N atoms are connected by H from both sides. Fig. 2 shows two ground-state structures with the lowest enthalpies (see Supplementary Fig. S2 for more other metastable phases, and Supplementary Table S1 for detailed structure information): the *P*-1 phase (found at 50 GPa) and the *P*$2_1$/c phase (found at 100 and 200 GPa). For the low-pressure *P*-1 phase, within the nearly planar polymeric nitrogen chains, H atoms bond with two adjacent N located at one side of armchair edge. Such configuration leads to the appearance of two types of N atoms — the one bonding with H (denoted as $N_H$) and the one not bonding with H (denoted as $N_N$), and three types of N-N bonds — the one between $N_H$ (denoted as $d_{HH}$), the one between $N_N$ (denoted as $d_{NN}$) and the one between $N_H$ and $N_N$ (denoted as $d_{NH}$). The lengths of these bonds are in the order of $d_{NN} < d_{NH} < d_{HH}$, with the value of 1.27, 1.32 and 1.36 Å, respectively at 50 GPa (see Supplementary Fig. S3 for the pressure dependence of them). By comparison with the N-N double-bond in the $HN_3$ molecule (1.23 Å)[51] and the single-bond in the polymeric cg-N (1.37 Å),[6] $d_{NN}$ can be reasonably classified as the N-N double-bond and $d_{HH}$, $d_{NH}$ as the single-bond. Such classification is also supported by the following electronic structure analysis. The bond between N and H (denoted as $d_{N-H}$, 1.04 Å at 50 GPa) is slightly larger than that of ammonia (1.01 Å) and shows an abnormal increase under compression (as shown in Supplementary Fig. S3; the underlying mechanism to be discussed later). For the high-pressure $P2_1/c$ phase, while the main features are similar to those of the *P*-1 phase, the H atoms are found to reside exactly on the connecting line between two N atoms from different chains. At high pressures this configuration gives rise to the strong hydrogen bonds between H and the N belonging to the adjacent chain (as denoted by dot lines in Fig. 2), which play an important role in stabilizing the structure (see below). As for the structure with H atoms on alternate N atoms (in the space group of *C*c), it is apparently not energetically favored compared with the *P*-1 and $P2_1/c$ structure in the lower-pressure region. The mechanism underlying its instability may be attributed to the appearance of only one type of N-N bond (i.e. $d_{NH}$ between $N_N$ and $N_H$) within the nitrogen chains, not satisfying the octet rule for N atoms. With increasing pressures, it shows a trend of being stabilized owing to the role played by the strong hydrogen bonds between H and the N belonging to the adjacent chain, similar to the case of the $P2_1/c$ structure.

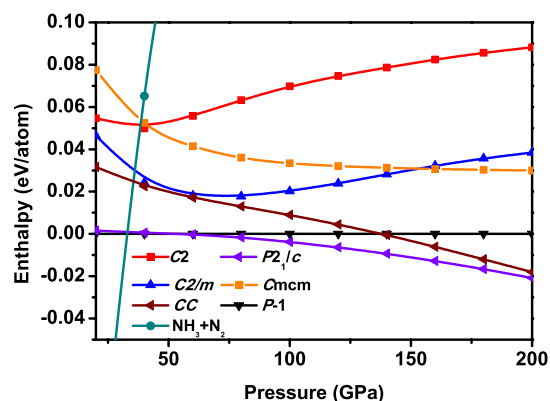

**Figure 3.** Calculated enthalpy (relative to the low-pressure ground-state *P*-1 phase) as a function of pressure for various structures of the $N_2H$ stoichiometry. The result of dissociation into $NH_3+N_2$ is also shown, and the actual phases at corresponding pressures of solid $NH_3$ (0 GPa: $P2_13$, 10-100 GPa: $P2_12_12_1$, above 100 GPa: $Pma2$) and $N_2$ (0 GPa: α-N, 10-50 GPa: ε-N, 60-180: cg-N, above 190 GPa: $Pba2$) are used for calculating the enthalpy of the mixture.

The enthalpy-pressure relationships of the ground-state $P$-1 and $P2_1/c$ phases together with other metastable phases are shown in Fig. 3, where the vdWs correction is included to properly consider the inter-chain interactions. The inclusion of the vdWs leads to dramatic enthalpy change at the low-pressure region for all the structures (see Supplementary Fig. S4 for the results excluding the vdWs). Generally the energy difference among these structures is not remarkable (< 0.1 eV/atom), which underlies that the phase stability is predominated by the framework of polymeric nitrogen chains. At low-pressure region, the enthalpy difference between the $P$-1 and $P2_1/c$ phases is almost indistinguishable, and the $P$-1 phase is slightly favored. Beyond the transition pressure of 55 GPa, the $P2_1/c$ phase dominates as the most stable structure. The stability of the $P2_1/c$ phase at high pressures can be rationalized in terms of the energy lowering by gradually increased strength of hydrogen bonds as the result of decreased distance between two polymeric nitrogen chains under compression. The decomposition pressure from the $P$-1 phase into $NH_3+N_2$ is ~33 GPa. We also explored the effect of zero-point energy (ZPE) correction (based on zone-center phonon modes), and the results indicate that the ZPE further shifts the decomposition pressure down to ~20 GPa and exhibits a minor effect on the transitions between different phases (see Supplementary Fig. S4).

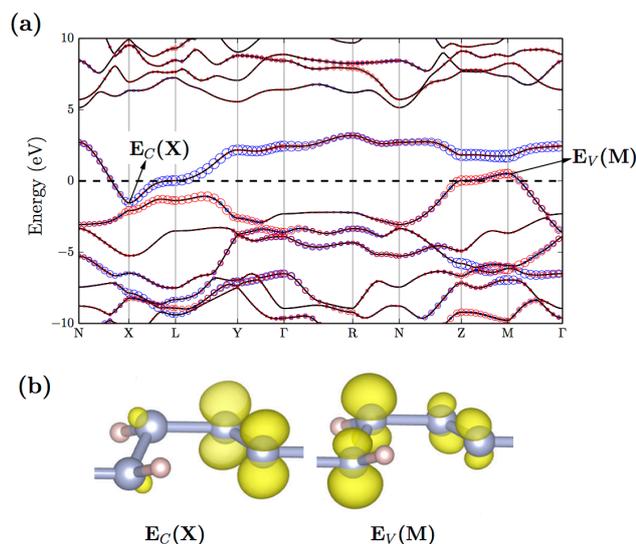

**Figure 4.** (a) Calculated band structure (with DFT) for the $P$-1 phase of $N_2H$ at 50 GPa. Projections onto the π orbital of $N_N$ and $N_H$ are shown in blue and red circles, respectively, and the projection magnitude is proportional to the circle size. The minimum of "unoccupied" band ($E_C(X)$) and the maximum of "occupied" band ($E_V(M)$) are indicated. (b) Band decomposed charge density of the $E_C(X)$ and $E_V(M)$ states.

**Novel metallicity caused by pressure-enhanced charge transfer of the polymeric $N_2H$ phases.** To our best knowledge, all of hydronitrogens discovered so far are insulating, as easily satisfying the octet rule usually prefers to open a band gap. This picture is also expected to be applicable to the $N_2H$ phases with polymeric nitrogen chains, where all of the N atoms participate in three bonds (by assigning $d_{NN}$ to the N-N double-bond and $d_{HH}$, $d_{NH}$ to single-bonds as above). Since the strong σ-bond and π-bond of $d_{NN}$ and N-H bond $d_{N-H}$ should create the bonding state in relatively high binding energy region, the band gap is thus to form between the highest occupied bonding state of the σ-bond of $d_{HH}$ (longer, and thus weaker than $d_{NH}$) and the lowest unoccupied anti-bonding states of the π-bond of $d_{NN}$. However, such speculation is only partially correct, as indicated by the band structure of the low-pressure $P$-1 phase in Fig. 4a: although the highest occupied and lowest unoccupied bands show expected bonding-orbital characteristics, a Fermi level is found to cross the lowest unoccupied band around X point (denoted as $E_C(X)$) and the highest occupied band around M point (denoted as $E_V(M)$), giving rise to a clear metallic feature. By projecting the band structure onto relevant bonding-orbitals, we find that the $E_C(X)$ and $E_V(M)$ states originate predominately from the π orbital of $N_N$ (represented by blue circles) and $N_H$ (represented by red circles), respectively, as also clearly shown in Fig. 4b. The charge redistribution associated with the occupancy of $E_C(X)$ and the unoccupancy of $E_V(M)$, leads to a charge of ~0.16 $e$ transferred from $N_H$ to $N_N$. Based on these analyses, we can rationalize the metallicity of the system in terms of two charge transfer processes: (i) the σ → π charge transfer within $N_H$; (ii) the charge transfer from the π-orbital of $N_H$ to the π-orbital of $N_N$. In the language of chemistry, these charge transfer processes can also be understood by using the concept of resonance effect occurring in the conjugated system having alternate σ and π bonds, where the induced delocalization of π electrons is responsible for the metallicity.[52] Similar metallic feature is also observed in the high-pressure $P2_1/c$ phase (see Supplementary Fig. S5 for its band structure at 50 GPa), where the metallicity can be rationalized in the same way.

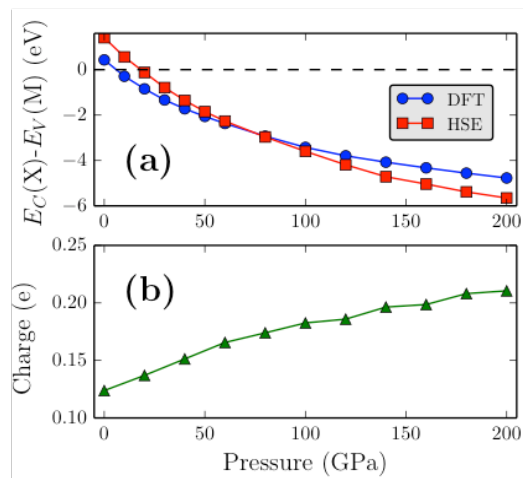

**Figure 5.** (a) Energy difference between $E_C(X)$ and $E_V(M)$ in Fig. 4 calculated with DFT and HSE, as well as (b) charge transferred from $N_H$ to $N_N$ (obtained from Bader charge analysis) as a functional of pressure for the $P$-1 phase of $N_2H$.

Fig. 5a shows the "degree of metallicity", i.e. the energy difference between $E_C(X)$ and $E_V(M)$ for the $P$-1 phase of $N_2H$ as the function of pressure, calculated with DFT, as well as the higher level HSE. The results clearly indicate that the metallicity is gradually induced by increasing pressure. With the well-known issue of underestimating band gap, DFT gets an insulator-to-metal transition at ~5 GPa, whereas HSE yields a higher transition pressure of ~20 GPa. Note that the more dramatic change of energy difference at high pressures rendered by HSE may imply its failure in properly treating the metallic system. The behavior of enhanced metallicity under compression corresponds well to the pressure dependence of the charge transferred from $N_H$ to $N_N$, as shown in Fig. 5b, where the charge transfer shows monotonic increase with increasing pressure. This provides clear evidence to the close relationship between the metallicity and charge transfer processes as addressed above. Because of this pressure-enhanced charge transfer from $N_H$ to $N_N$, the N-H bond ($d_{N-H}$) anchored to $N_H$ is weakened, which explains the abnormal increase of $d_{N-H}$ with pressure as shown in Supplementary Fig. S3.

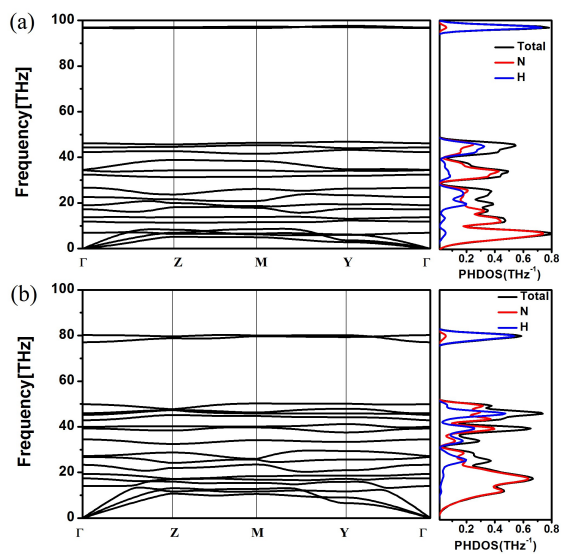

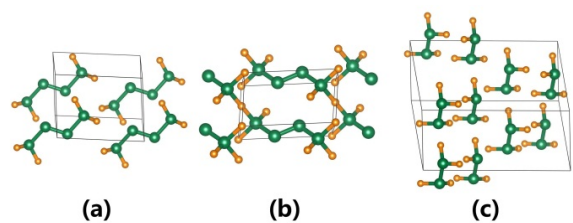

**Figure 7.** Stable crystal structures of the NH stoichiometry for (a) the low-pressure phase in the *C*2 space group and (b) the high-pressure *P*-1 phase with hydrogen-bond symmetrisation, and (c) of the $NH_2$ stoichiometry in the *Cc* symmetry. See Supplementary Table S1 for detailed structure information.

**Stable structures of other relatively nitrogen-poor stoichiometries.** As to the marginally stable NH stoichiometry (as in Fig. 1), at the pressures of 50 and 100 GPa our structure search found a *C*2 phase (Fig. 7a), which consists of similar tetrazene ($N_4H_4$) molecules to that of Ref. 20, but arranged in different spatial configuration. With the pressure increasing to 200 GPa, another phase in the *P*-1 symmetry (Fig. 7b) becomes energetically favored. This phase is also composed of $N_4H_4$ molecules, which are connected together by the bridge H atoms via hydrogen-bond symmetrisation, resembling the case of ice-X.[53] In any case, the previously proposed polymeric $P2_1/m$ structure consisting of the infinite zigzag nitrogen chain[27] is never identified by our structure search up to 200 GPa. Comparing with the above predicted stable phases, it demonstrates a notable energetical instability by ~130 meV/atom at 75 GPa and ~158 meV/atom at 200 GPa. Another stoichiometry that may be stabilized is $NH_2$ at 200 GPa, which adopts a structure with the *Cc* space group, made up of hydrazine ($N_2H_4$) molecules.[17] These results indicate that in the relatively nitrogen-poor region (N:H ≤ 1), the stable structures of hydronitrogens are always dominated by the molecular phases.

**Figure 6.** Calculated phonon band dispersion and projected density of states (onto individual atomic species) for the *P*-1 phase of $N_2H$ at 0 (a) and 50 (b) GPa.

**Lattice dynamical stability of the polymeric $N_2H$ phases.** One of important criterions to judge whether new material is stable or not is the stability of lattice dynamics, which requires no imaginary mode appearing in the phonon spectrum in the whole Brillouin zone. Fig. 6 shows the phonon band dispersion and density of states for the *P*-1 phase of $N_2H$ (see Supplementary Fig. S6 for the phonon spectrum of the high-pressure $P2_1/c$ phase). The lattice dynamical stability down to 0 GPa is clearly evidenced by the absence of any imaginary phonon mode. It thus allows for the possibility of stabilizing the polymeric $N_2H$ phases under some kinetic regime at ambient conditions (*e.g.* by rapid quenching from high temperature). Note that the hydrogen stretching mode, which corresponds to the strong blue peak located at the high-frequency region in the projected density of states, decrease from ~100 THz at 0 GPa to ~80 THz at 50 GPa. This unusual red shift of phonon frequencies under compression is in accord well with the abnormal increase of $d_{N-H}$ with pressure caused by pressure-enhanced charge transfer from $N_H$ to $N_N$ as discussed above.

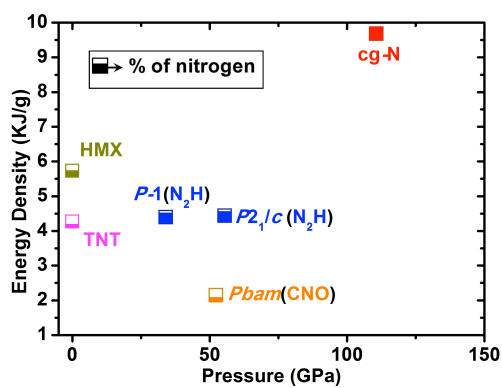

**Figure 8.** Calculated energy densities of the polymeric $N_2H$ phases at the ambient pressure in relation to the pressure required to stabilize them. The energy densities of cg-N,[5,6] CNO,[13] as well as the standard energetic materials (HMX and TNT)[54] are also shown for comparison. The nitrogen content is denoted by the filled area of squares. For HMX and TNT, the common values of heat of denotation[54] are taken.

**Energy densities of the polymeric N₂H phases.** The predicted polymeric N₂H phases, in which the infinite armchair-like nitrogen chains contains alternate N-N single-bonds and double-bonds, are expected to be favorable for HEDM. In Fig. 8, we show their energy densities at the ambient pressure in relation to the pressure required to stabilize them. The energy density is evaluated by assuming that (i) the polymeric N₂H phases can be stabilized down to the ambient pressure (ii) the final products after decomposition are NH₃ and N₂, *i.e.*, $N_xH_y \rightarrow (x-y/3)/2 \, N_2 + y/3 \, NH_3$. The resulted energy densities are 4.40 and 4.44 KJ/g for the *P*-1 and *P*2₁/c phase, respectively. The values are, though still much lower than 9.7 KJ/g of cg-N owing to the presence of N-N double-bonds and N-H bonds, two times larger than 2.2 KJ/g of the CO/N₂ mixture (CNO)[13]. Also the content of nitrogen in N₂H (96.6%) is obviously much higher than that of CNO. Moreover, the stabilization pressure for the *P*-1 phase of N₂H is only ~33 GPa, much lower than the pressure required for stabilizing the polymeric *cg*-N (~94 GPa/127 GPa) as mentioned. Note that the energy densities of the polymeric N₂H phases are mildly lower than those of the standard energetic materials, TNT (4.3 kJ/g) and HMX (5.7 kJ/g),[54] which are certainly thermodynamically stable at ambient conditions.

## IV. Conclusions

To conclude, by combining a crystal structure search method and first-principle total energy minimization calculations, we explore the energy landscape of hydronitrogen system at high pressures to seek for the non-molecular phase of polymeric nitrogen for high energy density applications. Different from the conventional approach for synthesizing hydronitrogens by mixing directly N₂ and H₂, we propose an alternative synthesis procedure to make hydronitrogens at high pressures via the reaction between N₂ and NH₃. This allows us to zoom in the region of nitrogen-rich stoichiometries and simultaneously consider the effect of competing phase (NH₃) on phase stabilities. We discover two novel stable phases of hydronitrogen in the N₂H stoichiometry (with the space group of *P*-1 and *P*2₁/c) that consist of the polymeric infinite armchair-like nitrogen chains. The N₂H phases are predicted to be stabilized against decomposition into N₂ and NH₃ above ~33 GPa. Different from other common hydronitrogens, they exhibit exotic metallic feature, which can be rationalized in terms of pressure-enhanced charge transfer within nitrogen chains and the delocalization of π electrons in the conjugated system. Due to the high energy densities (4.40 and 4.44 KJ/g for the *P*-1 and *P*2₁/c phase, respectively) together with the low stabilization pressure, the predicted N₂H phases feature as possible candidates for high energy density materials. It should be pointed out that even though the polymeric N₂H phases show lattice dynamical stability, allowing for the possibility of kinetic stability, they are still thermodynamically unstable against decomposition at ambient pressure. Hence, to realize their practical application, much more experimental efforts are required to verify their stability at high pressures and attempt to stabilize them under some kinetic regime at ambient conditions.

## Acknowledgements

We are grateful to John Tse and Yixin Zhao for helpful discussion. This work is supported by the Recruitment Program of Global Experts (the Thousand Young Talents Plan), the National Natural Science Foundation of China (under Grants No. 11404128), the Postdoc-toral Science Foundation of China (under Grant No. 2014M551181) and the Startup Funding of Jilin University. We acknowledge the use of computing facilities at High Performance Computing Center of Jilin University.

## Notes and references

[a] State Key Laboratory of Superhard Materials, Jilin University, Changchun 130012, China

[b] College of Materials Science and Engineering, Jilin University, Changchun 130012, China

[*] Address correspondence to: wyc@calypso.cn or lijun_zhang@jlu.edu.cn

†Electronic Supplementary Information (ESI) available: [Detailed description on crystal structure search procedures, structural information for all the energetically stable phases of hydronitrogen, as well as pressure dependence of bond lengths, band structure and phonon band dispersion for some hydronitrogen phases]. See DOI: 10.1039/b000000x/

# Supplementary Methods

To seek for energetically stable hydronitrogens, we perform a global minimization of the energy landscape of N-H system at high pressures by combining *ab initio* density functional theory (DFT) total-energy calculations with a particle swarm optimization (PSO) algorithm as implemented in the CALYPSO code. Structure searches are performed at 50, 100, and 200 GPa with up to eight formula units for the hydronitrogens with varied H and N contents. Firstly, a population of random structures with certain crystallographic symmetry are constructed (as the first generation), in which the internal atomic positions are generated by symmetry operations of randomly selected space groups. Then the structures are optimized to the free energy local minima with DFT calculations. By evaluating the enthalpies of these structures, 60% of them with the lower enthalpies, together with the 40% newly generated structures, are used to produce the structures of next generation by the structure operators of PSO. In this step, a structure fingerprinting technique of bond characterization matrix is applied to generate new structures, so that identical (or extremely similar) structures are strictly forbidden. These significantly enhance the diversity of the generated structures, which is crucial for final convergence of the global structure search. The local structure optimizations are performed by using the conjugate gradients method. For most of the cases, the structure searches reach the convergence (*i.e.*, no new structure with the lower energy emerging) after 50 generations, at which more than thousand structures are evaluated.

# Supplementary Tables

**Supplementary Table S1.** Detailed structural information for the energetically stable phases of hydronitrogen identified in our structure search.

| Chemical composition | Space group | Pressure (GPa) | Lattice parameters (Å) | Atomic coordinates (Fractional) |
|---|---|---|---|---|
| $N_2H$ | $P$-1 | 50 | a=3.689 b=3.547 c=2.677 α=71.35 β=78.67 γ=66.44 | N1 2i 0.6750 0.3692 0.1068<br>N2 2i 0.6745 0.9942 0.4148<br>H1 2i 0.0775 0.6700 0.1662 |
| $N_2H$ | $P2_1/c$ | 200 | a=2.354 b=6.093 c=3.137 α=90.00 β=104.66 γ=90.00 | N1 4e 0.5982 0.5963 0.0536<br>N2 4e 0.9291 0.9036 0.9438<br>H1 4e 0.7119 0.2340 0.5011 |
| $N_4H_4$ | $C2$ | 100 | a=3.396 b=3.977 c= 4.498 α=90.00 β=93.82 γ=90.00 | N1 2c 0.7495 0.6545 0.6737<br>N1 2c 0.9473 0.6605 0.1288<br>H1 2c 0.9853 0.1611 0.7797<br>H2 2c 0.2590 0.3656 0.5227 |
| $N_4H_4$ | $P$-1 | 200 | a=4.276 b=2.469 c=2.424 α=98.55 β=87.52 γ=95.62 | N1 2i 0.8289 0.2375 0.7746<br>N2 2i 0.3712 0.0906 0.5132<br>H1 1a 0.0000 0.0000 0.0000<br>H2 2i 0.2852 0.4832 0.9316<br>H3 1g 0.0000 0.5000 0.5000 |
| $NH_2$ | $CC$ | 200 | a=3.437 b=8.056 c=2.527 α=90.00 β=116.84 γ=90.00 | N1 2a 0.0930 0.8654 0.1244<br>N2 2a 0.9876 0.6334 0.6464<br>H1 2a 0.0205 0.4607 0.9160<br>H2 2a 0.5590 0.4579 0.3583<br>H3 2a 0.0338 0.7420 0.4041<br>H4 2a 0.3153 0.6353 0.1721 |

# Supplementary Figures

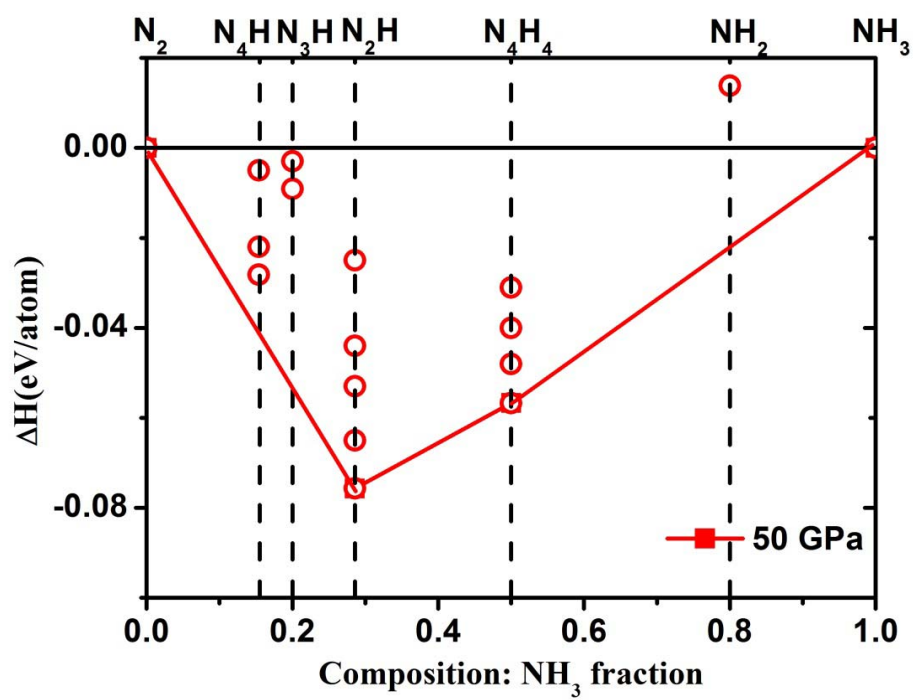

**Supplementary Figure S1.** The convex hull of the N-H system with respect to $NH_3$ and $N_2$ as the binary variables at 50 GPa.

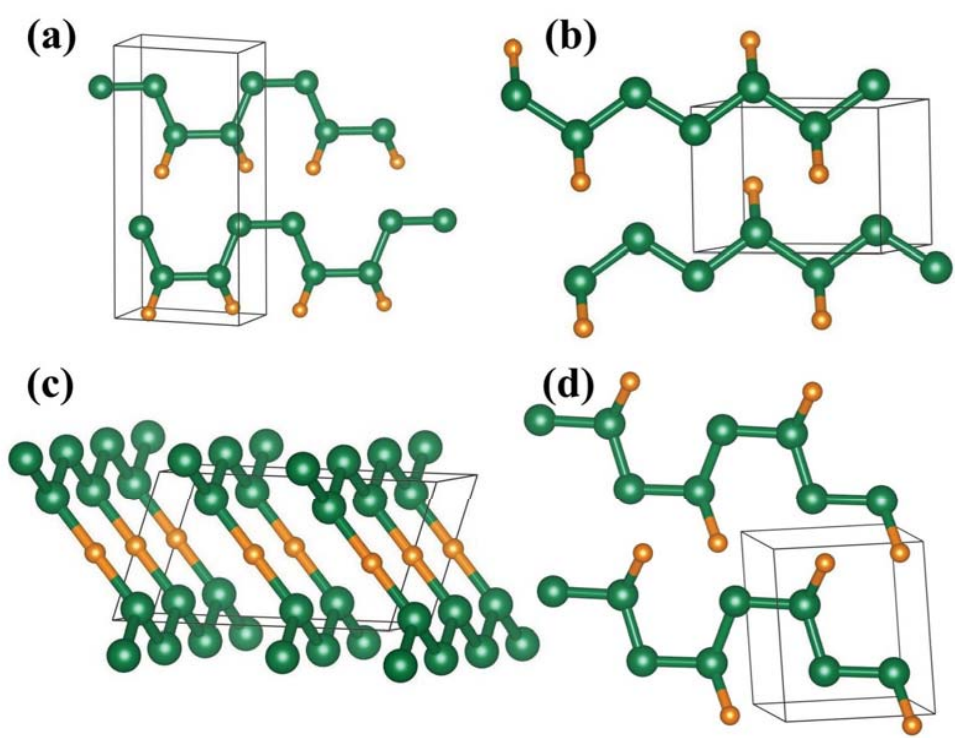

**Supplementary Figure S2.** Crystal structures of the N₂H polymeric phase with (a) the *C*2 space group, (b) the *C*2/m space group, (c) *C*mcm space group, and (d) *C*c space group.

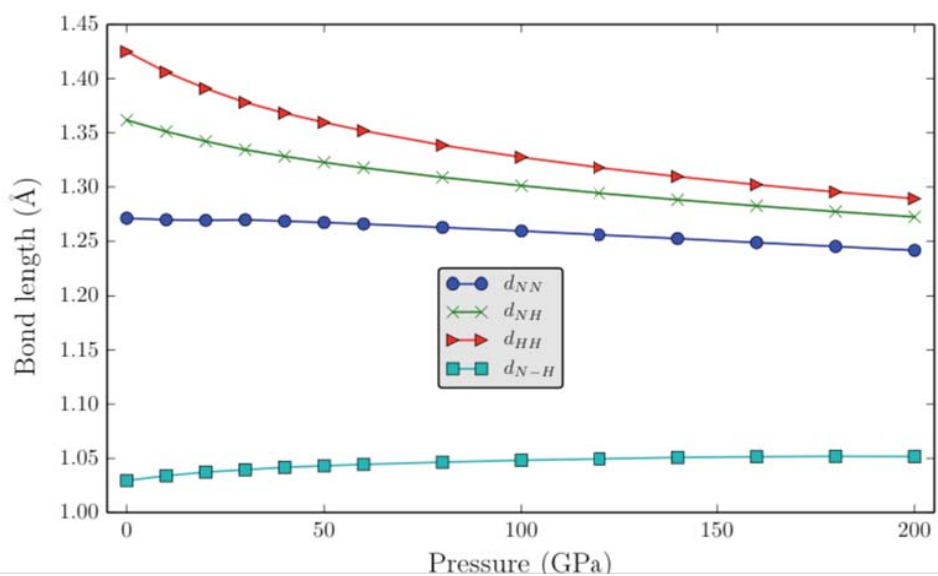

**Supplementary Figure S3.** Pressure dependence of bond lengths of the bond between $N_H$ (denoted as $d_{HH}$), the one between $N_N$ (denoted as $d_{NN}$), the one between $N_H$ and $N_N$ (denoted as $d_{NH}$), and the one between N and H (denoted as $d_{N-H}$) for the low-pressure $P$-1 phase of $N_2H$.

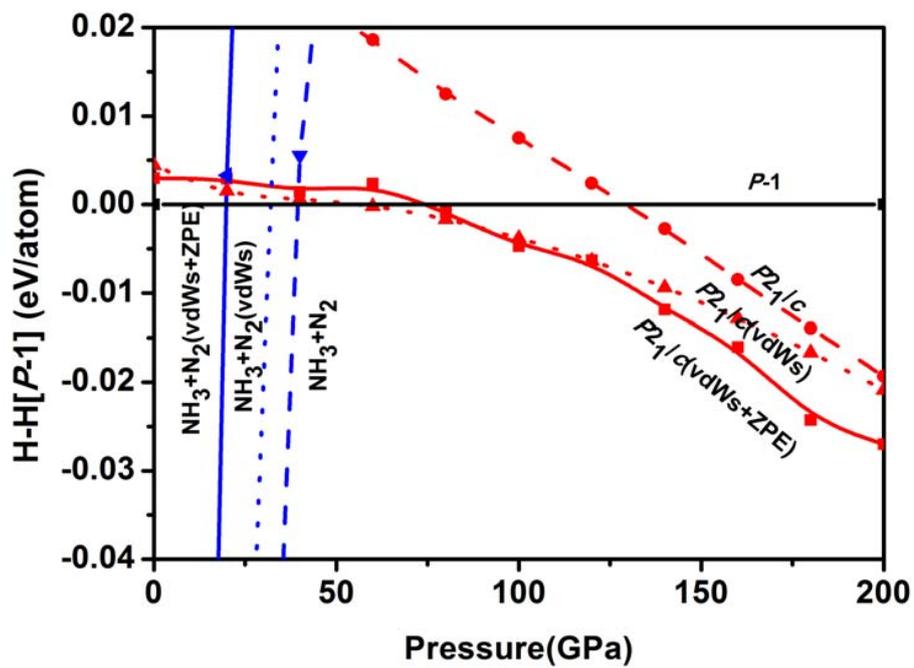

**Supplementary Figure S4.** Comparison of the enthalpy-pressure relationship by using different levels of approach (DFT, DFT+vdWs, DFT+vdWs+ZPE) for the stable low-pressure *P*-1 phase, high-pressure $P2_1/c$ phase, as well as dissociation into $NH_3+N_2$. The energy is in eV per atom and relative to the *P*-1 phase.

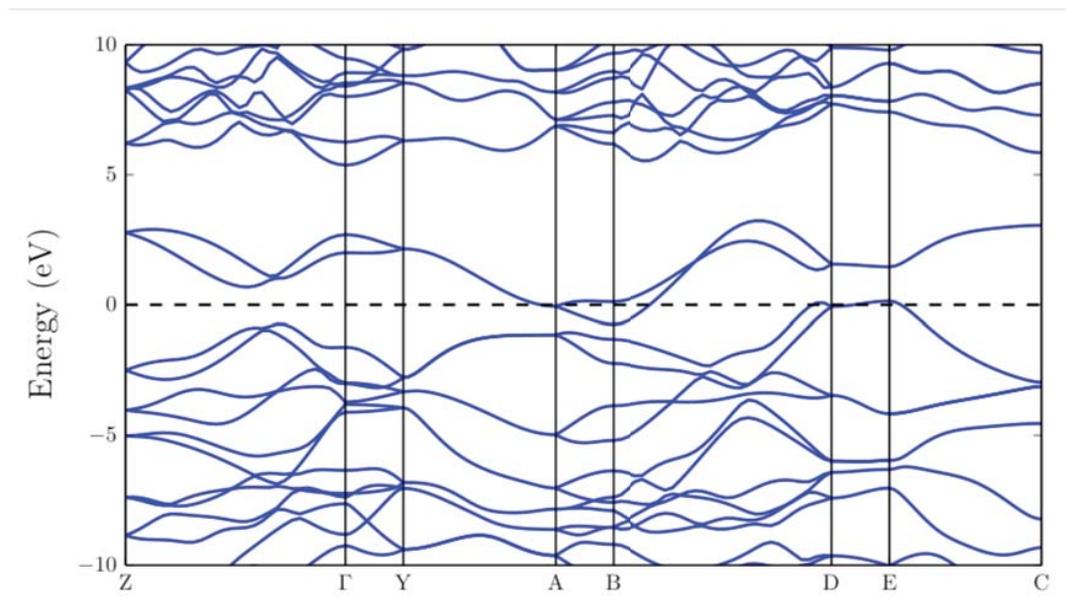

**Supplementary Figure S5.** Calculated band structure (with DFT) for the high-pressure $P2_1/c$ phase of $N_2H$ at 50 GPa.

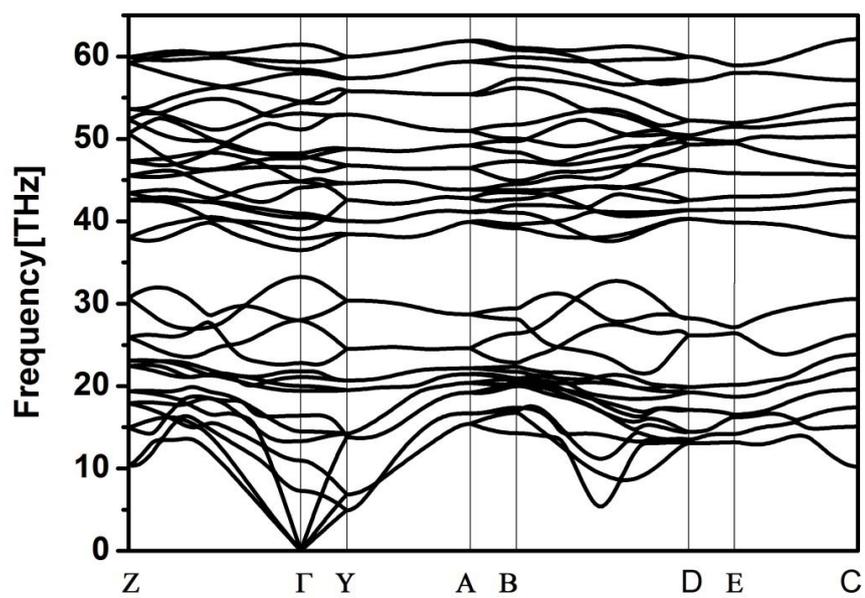

**Supplementary Figure S6.** Calculated phonon band dispersion for the high-pressure $P2_1/c$ phase of $N_2H$ at 200 GPa.